# Lateral Heterojunction $Sb_2Te_3$/$Bi_2Te_3$ and its topological transport

Fucong Fei[1], Qianjin Wang[2], Zhongxia Wei[1], Rui Wang[1], Danfeng Pan[1], Bo Zhao[1], Xuefeng Wang[3], Xinran Wang[3], Jianguo Wan[1], Fengqi Song[1,*], Baigeng Wang[1,*], Guanghou Wang[1]

[1]National Laboratory of Solid State Microstructures, Collaborative Innovation Center of Advanced Microstructures, and College of Physics, Nanjing University, Nanjing, 210093, P. R. China

[2]National Laboratory of Solid State Microstructures, Collaborative Innovation Center of Advanced Microstructures, and Department of Material Science and Engineering, Nanjing University, Nanjing, 210093, P. R. China

[3]National Laboratory of Solid State Microstructures, Collaborative Innovation Center of Advanced Microstructures, and School of Electronic Science and Engineering, Nanjing University, Nanjing, 210093, P. R. China

---

[*] Corresponding authors. F. S. (songfengqi@nju.edu.cn) and B. W. (bgwang@nju.edu.cn), Fax +86-25-83595535

**Abstract**

A lateral heterojunction of topological insulator $Sb_2Te_3$/$Bi_2Te_3$ was successfully synthesized using a two-step solvothermal method. The two crystalline components were separated well by a sharp lattice-matched interface when the optimized procedure was used. Inspecting the heterojunction using high-resolution transmission electron microscopy showed that epitaxial growth occurred along the horizontal plane. The semiconducting temperature-resistance curve and crossjunction rectification were observed, which reveal a staggered-gap lateral heterojunction with a small junction voltage. Quantum correction from the weak antilocalization reveals the well-maintained transport of the topological surface state. This is appealing for a platform for spin filters and one-dimensional topological interface states.

**Introduction**

Lateral heterojunctions (LHJs) play a more interesting role than vertical ones in the heterojunction physics of topological insulators (TIs) because in a vertical TI heterojunction a surface is eliminated, killing a surface state (SS), whereas a new interface state is generated between two well-maintained SS regions in the same plane in a TI-based LHJ[1-3,4]. Such interface states may accommodate novel electronic physics for Dirac fermions[1,2,4,5]. For example, aligning the working functions of the two crystals in a LHJ will tune the eigenvalues of the spin helicity of the SS between +1 and −1, allowing the LHJ to be used in quantum information devices[1,2]. Spintronic devices can also be implemented in such a TI-based LHJ. It has been proposed that quantum spin Hall edge states will arise and accumulate spins according to the distinct behaviors of the two TIs when a Landau quantization magnetic field is applied[1,2]. A spin-filtered state will therefore be expected along the interface. The ability to control the SS and the bulk electrons separately also points to some new physics of LHJs. LHJs also have an advantage of tunable rectification than the traditional three-dimensional heterojunctions[6] It has been proposed to implement such idea by fractional gating [1], but no experiments have yet been described in which the growth of such a TI-based LHJ has been achieved. Recent progress in the production of LHJs in two-dimensional materials has shed light on this topic[6,7,8].

Here we report the successful preparation of $Sb_2Te_3$/$Bi_2Te_3$ LHJs. $Sb_2Te_3$ and $Bi_2Te_3$ are both TI materials with 5–20 quintuple layers[9,10]. High-resolution (HR) transmission electron microscopy (TEM) images of the LHJ interface showed that epitaxial growth occurred. Electrical rectification was observed in the cross-junction transport in the LHJ devices, indicating the onset of an interior electric field across the junction interface. Weak antilocalization and its robust coupling were found, showing that topological transport was well maintained.

**Synthesis and Characterization**

All of the LHJ flakes used in this study were prepared using a simple two-step solvothermal technique. The first step was preparing $Bi_2Te_3$ nanoflakes[11]. Powdered $Bi_2O_3$ (0.466g), $TeO_2$ (0.48g), and NaOH (1g) were dissolved in ethylene glycol (50ml) containing polyvinylpyrrolidone (PVP) surfactant (0.8g). The solution was annealed at 180 ℃ in an autoclave for 12 h. The second step was introducing 8ml of the first product to the coating solution, which was prepared by dissolving powdered $SbCl_3$ (0.228g), $TeO_2$ (0.24g), and NaOH (0.9g) in a mixture of ethylene glycol (38ml) and deionized water (4ml) containing PVP (0.8g). The second step was completed by annealing the mixture at 195 ℃ in an autoclave for 24 h. Core-shell structures, rather than LHJs, were produced in the absence of deionized water in the second step. The products were examined by HRTEM using a FEI TECNAI F20 transmission electron microscope equipped with a 200kV electron gun and an energy dispersion spectrometer. Atomic force microscopy images were acquired using an NT-MDT atomic force microscope. An ARL X'TRA X-ray powder diffractometer was used to determine the crystal structure. Coarse electrodes were fabricated using ultraviolet lithography technique and the focused ion beam deposited tiny Pt contacts were fabricated by a FEI Helios Nanolab 600i dual beam system. Electrical transport measurements were performed in a Cryomagnetics cryostat equipped with a Stanford Research Systems SR830 digital lock-in amplifier, Keithley 4200 and Agilent 2635 meters, and a Quantum Design PPMS-16T system. A temperature of 2K and a magnetic field of 15 T were achieved during the measurements.

**Preparation of the LHJ using the two-step route**

The first step of the synthesis produced $Bi_2Te_3$ nanoflakes[11]. These nanoflakes were then dispersed in the coating solution that was intended to prepare $Sb_2Te_3$ nanoflakes. Both these crystals have been shown to be TIs when they are more than six quintuple layers (nanometers) thick [9]. These crystals were selected so that lateral expitaxy could be achieved, because the crystals have similar space groups and lattice constants.

The successfully synthesized TI LHJs were $Bi_2Te_3$ nanoplates surrounded by $Sb_2Te_3$ margins. Low magnification TEM images of the material are shown in **Figure 1(a)**, in which regular hexagonal shapes and shiny fringes (from electronic interference) can be seen, indicating a good degree of crystallinity. The inner plates can be identified by the pale contrast in **Figure 1(a)**. The contrast was enhanced by acquiring a scanning TEM image using the high-angle angular dark-field mode, in which atomic-weight-dependent electronic scattering leads to intense atomic contrast. The inner flakes can be seen clearly in **Figure 1(b)** since the heavier elements (Bi) were more enriched in the inner crystals than the marginal ones. The X-ray diffraction pattern of the prepared sample (dry powder) is shown in **Figure 1(c)**, and all the peaks were assigned according to the reference spectrum (the lower panel; JCPDS cards

15-0874 and 15-0863). The peaks were therefore found to be for $Sb_2Te_3$ and $Bi_2Te_3$, which were the intended products. The strong peaks of two candidates were resolved well, indicating the nice phase separation. The energy dispersive spectra for a single LHJ flake are shown in **Figure 1(d)** (the flake is shown in the inset). The two regions were respectively found to be composed of $Bi_2Te_3$ and $Sb_2Te_3$. We are therefore convinced that the desired LHJ was successfully prepared using the two-step route shown schematically in **Figure 1(e)**.

**Lateral epitaxial growth between the two TIs**

Atomic force microscopy images of the flakes were acquired and inspected carefully. There was a hexagonal interface between the two regions in each flake, as shown in **Figure 2(a)**, and the interface was 1nm thicker than the other areas of the flake. We believe that this was because the chemically active interface absorbed some mobile contaminants[12]. The inner crystal and the margin crystal were both 16nm thick, as shown by the scanning profile along the blue line. This excludes the possibility that $Sb_2Te_3$ formed a coating on the top surface of the $Bi_2Te_3$. This was also confirmed because no obvious cross-doping was found in energy dispersive spectroscopy imaging in **Figure 1(d)**. In fact, $Bi_2Te_3/Sb_2Te_3$ core-shell structures were formed when water was excluded from the mixture used to perform the second step of the synthesis and when larger amounts of $TeO_2$ and $SbCl_3$ were used. Such well-limited lateral growth can be attributed to the guiding behavior of the surfactant template. It can also be attributed to the fact that the surface energy can be reduced during the lateral growth while less surface energy is saved in the Van der Waals epitaxial growth along vertical direction[8].

The HRTEM images showed that a good standard of epitaxial growth occurred. HRTEM images of positions near the LHJ flake interface are shown in **Figure 2(b)**, and parallel lattice fringes can be seen in both the inner and marginal crystals. Different lattice spacings (0.22nm and 0.214(3) nm) were found in the two regions of the LHJ, as shown in **Figure 2(c)** and **2(d)**, despite their having similar hexagonal fast Fourier transformation (FFT) patterns (shown in the insets). These spacings were respectively found to be for the $(11\bar{2}0)$ planes of the two crystals[10]. The similar lattice spacings allowed fine epitaxial growth between the two crystals as seen. Higher magnification images along the LHJ interface (**Figure 2(e)**) show that the two sets of atomic fringes evolved smoothly from left to right. The junction interface was sharp at the nanometer scale. The lattice mismatch has been solved by a dislocation, which is finally found as shown in the figure and marked by the arrow. It is therefore clear that lateral epitaxial growth occurred and that the LHJ had a lattice-matched abrupt interface.

**Electrical transport of the device**

We fabricated a two-probe device across the LHJ, as shown schematically in the inset of **Figure 3(a)**. The as-prepared flakes were generally covered with a thin layer of organic surfactant, which protected the flakes from being oxidized but also prevented electrical transport. The presence of such a surfactant layer always causes insulating failure of the devices. We made use of the insulating organic coating while the device was being fabricated using a focused ion beam. We tentatively dig the LHJ surface and remove the organic layer for 5nm in the selected region with a diameter of 100nm, leaving the surface of the sample bare to allow a contact to be fabricated. Pt electrodes were prepared on the bare areas, using a Pt nanogun, to complete the device. This produced flying leads across the junction. Perfectly linear current–voltage (IV) curves were found, at least at room temperature, indicating that successful Ohmic contacts had been produced. The temperature dependence of the resistance of a typical device is shown in **Figure 3(a)**, showing that the resistance increased from 100 kΩ to around 600 kΩ as the temperature decreased. This indicates that the device behaved as a semiconductor and that the Fermi level was aligned with the bandgap.

The signature of cross-junction transport was found at low temperatures. The linear current–voltage curve changed as the temperature decreased. A rectification effect occurred at 5 K, as shown in **Figure 3(b)**. The transport current quickly increased to 5nA when the positive voltage was 2mV, but the transport current remained well below 1nA when a negative bias voltage was applied. This is a typical rectification effect, indicating the onset of an interior electrical field across the LHJ interface. The current increased abruptly and quickly at −3mV when the voltage was decreased, indicating that the threshold voltage was small, ~3mV. This threshold voltage is very small, and we concluded that it was because $Bi_2Te_3$ and $Sb_2Te_3$ have similar band structures. We looked for temperature-dependent current–voltage asymmetry for the interior field in the junction and found that voltage asymmetry was greatly reduced from voltages of 3mV with the increasing temperature, and disappeared when the temperature was higher than 30K, as is shown in **Figure 3(c)**. Taking the Boltzmann constant (~2.5meV at 30K) into consideration, our observations confirm that there was a small but robust junction voltage of ~3mV.

We propose that the LHJ produced was a heterostructure with a staggered bandgap, as shown in **Figure 3(e)**. The LHJ was an n–n junction. Charge transfer experiments using ethylenediamine tetraacetic acid (EDTA) molecules showed that the Fermi level was near the conduction band because the EDTA molecule often introduce n-type carriers and its introduction was found to decrease the resistance of the sample. The Fermi level was quite near the bottom of the conduction band (5–10meV) in the bulk gap of ~160meV, as determined from the thermally activated conductivity. The interior voltage drop across the junction was found to be around 3meV, and the width of the junction region was subject to the carrier diffusion constants, which were not determined. The carriers' density was higher than $10^{12} cm^{-2}$, which are mostly surface carriers and can't be controlled yet by a SiO2 back gate[13]. The system still requires delicate optimization to allow a p–n junction to be formed between the two SS regions, or intense gate of $SrTiO_3$[14].

The topological SS are found well maintained after the magnetoresistance (MR) of the device was measured at 2 K and at up to a field of 15 T. A crossover was seen from a low-field parabolic MR to an increasing high-field linear MR, as shown in **Figure 3(d)**. Such a trend has been found widely in Dirac systems, in which low-field MR has been attributed to classical MR and high-field linearity has been attributed to a high g factor or quasilinear dispersion[15]. The classical parabolic MR gives the carriers a mobility of a few hundred square centimeters per volt and a carrier concentration of more than $7\times10^{12}\text{cm}^{-2}$. An abrupt decrease in the resistance was found near the zero field, and this was attributed to weak antilocalization, which has recently been regarded to be a signature of SS transport[13,16]. We fitted a line to the zero-field MR curve using the Hikami–Larkin–Nagaoka formula as shown in the inset of Figure 3(d) [17,18], and this gave a dephasing length of 97nm and an alpha value of 0.47. This alpha value indicates that the device had a single transport channel or electronic state. This can be understood because all SSs are coherently coupled because their electronic dephasing lengths are much larger than their thicknesses[19]. Our device had a dephasing length close to those obtained in samples without a junction. All these observations point to the transport through the topological SS.

In summary, a two-step solvothermal method was used to successfully synthesize LHJ samples based on $Bi_2Te_3$ and $Sb_2Te_3$ TIs. The LHJ had a lattice-matched abrupt interface and an interior crossjunction voltage of 3mV. The LHJ could be further optimized because the junction voltage drop was small and the Fermi level was inappropriate (because the two materials have similar band structures). These further improvements could result in the successful use of the LHJ in spin filters and IS devices.

**Acknowledgements**

The authors would like to thank the National Key Projects for Basic Research in China (Grant Nos. 2013CB922103, 2011CB922103, and 2014CB921103), the National Natural Science Foundation of China (Grant Nos. 91421109, 11023002,

11134005, 61176088, and 2117109), the NSF of Jiangsu Province (Grant Nos. BK20130054 and BC2013118), the PAPD project, and the Fundamental Research Funds for the Central Universities for financially supporting this work. Technical assistance provided by Prof. Li Pi and Mingliang Tian from the Hefei High Field Center is also gratefully acknowledged.

**References&Notes**

1. J. Wang, X. Chen, B. F. Zhu, and S. C. Zhang, Phys. Rev. B **85**, 235131 (2012).
2. R. Ilan, F. d Juan, and J. E. Moore, ArXiv, 1410.5823 (2014).
3. S. Sakhi, Europhys Lett **73** (2), 267 (2006); R. Fazio and H. van der Zant, Phys Rep **355** (4), 235 (2001); J. H. Bardarson and J. E. Moore, Rep. Prog. Phys. **76** (5), 056501 (2013); M. Z. Hasan and C. L. Kane, Rev. Mod. Phys. **82**, 3045 (2010).
4. A. R. Akhmerov, J. Nilsson, and C. W. J. Beenakker, Phys Rev Lett **102** (21) (2009); B. Doucot, M. V. Feigel'man, and L. B. Ioffe, Phys Rev Lett **90** (10) (2003); L. Fu and C. L. Kane, Phys Rev Lett **100** (9) (2008).
5. L. Fu and C. L. Kane, Phys Rev Lett **102** (21) (2009); X. Y. Lee, H. W. Jiang, and W. J. Schaff, Phys Rev Lett **83** (18), 3701 (1999); S. Mondal, D. Sen, K. Sengupta, and R. Shankar, Phys Rev Lett **104** (4) (2010); Y. Tanaka, T. Yokoyama, and N. Nagaosa, Phys Rev Lett **103** (10) (2009).
6. X. Hong, J. Kim, S. Shi, Y. Zhang, C. Jin, Y. Sun, S. Tongay, J. Wu, Y. Zhang, and F. Wang, Nature Nano **9**, 682 (2014).
7. C. Huang, S. Wu, A. Sanchez, J. Peters, R. Beanland, J. Ross, P. Rivera, W. Yao, D. Cobden, and X. Xu, Nature Materials **13**, 1096 (2014); X. Duan, C. Wang, J. Shaw, R. Cheng, Y. Chen, H. Li, X. Wu, Y. Tang, Q. Zhang, A. Pan, J. Jiang, R. Yu, Y. Huang, and X. Duan, Nature Nano **9**, 1024 (2014); M. P. Levendorf, C. Kim, L. Brown, P. Huang, R. Havener, D. A. Muller, and J Park, Nature **488**, 627 (2012); Z. Zheng Liu, L. Ma, G. Shi, W. Zhou, Y. Gong, S. Lei, X. Yang, J. Zhang, J. Yu, K. P. Hackenberg, A. Babakhani, J. Idrobo, R. Vajtai, J. Lou, and P. M. Ajayan, nature Nano **8**, 119 (2013); K. Yan, D. Wu, H. Peng, L. Jin, Q. Fu, X. Bao, and Z. F. Liu, Nature Communications **3**, 1280 (2012).
8. F. Withers, O. Pozo-Zamudio, A. Mishchenko, A. Rooney, A. Gholinia, K. Watanabe, T. Taniguchi, S. Haigh, A. Geim, A. Tartakovskii, and K. Novoselov, Nature Materials **Advance online**, 4205 (2015).
9. D. Hsieh, Y. Xia, D. Qian, L. Wray, F. Meier, J. H. Dil, J. Osterwalder, L. Patthey, A. V. Fedorov, H. Lin, A. Bansil, D. Grauer, Y. S. Hor, R. J. Cava, and M. Z. Hasan, Phys Rev Lett **103** (14) (2009); Guang Wang, Xie-Gang Zhu, Yi-Yang Sun, Yao-Yi Li, Tong Zhang, Jing Wen, Xi Chen, Ke He, Li-Li Wang, Xu-Cun Ma, Jin-Feng Jia, Shengbai Zhang, and Qi-Kun Xue, Advanced Materials **23** (26), 2929 (2011).
10. Haijun Zhang, Chao-Xing Liu, Xiao-Liang Qi, Xi Dai, Zhong Fang, and Shou-Cheng Zhang, Nat Phys **5**, 438 (2009).
11. D. Kong, K. J. Koski, J. J. Cha, S. S. Hong, and Y. Cui, Nano Lett. **13** (2), 632 (2013).
12. P. Jensen, Rev. Mod. Phys. **71** (5), 1695 (1999).
13. T. Chen, Q. Chen, K. Schouteden, W. K. Huang, X. F. Wang, Z. Li, Miao. F., X. R. Wang, Z. Li, B. Zhao, S. C. Li, F. Q. Song, J. L. Wang, B. G. Wang, C. Haesendonck, and G. H. Wang, Nature Communications **5**, 5022 (2014).
14. Guanhua Zhang, Huajun Qin, Jun Chen, Xiaoyue He, Li Lu, Yongqing Li, and Kehui Wu, Advanced Functional Materials **21** (12), 2151 (2011).
15. W. Wang, Y. Du, G. Xu, X. Zhang, E. Liu, Z. Liu, Y. Shi, J. Chen, C. Wu, and X. X. Zhang, Sci. Rep. **3**, 2181 (2013); K. Huynh, Y. Tanabe, and K. Tanigaki, Phys. Rev. Lett. **106**, 217004 (2011); T. Liang, Q. Gibson, M. Ali, M Liu, R. Cava, and N. P. Ong, Nature Materials, ADVANCE

ONLINE PUBLICATION (2014); H. Tang, D. Liang, R. Qiu, and X. Gao, ArXiv, 1101.2152 (2011); B. Assaf, T. Cardinal, P. Wei, F. Katmis, J. F. Moodera, and D. Heiman, Appl. Phys. Lett. **102**, 012102 (2013); X. Wang, Y. Du, S. Dou, and C. Zhang, Phys. Rev. Lett. **108**, 266806 (2012); B. Gao, P. Gehring, M. Burghard, and K. Kern, Appl. Phys. Lett. **100**, 212402 (2012).

16  H. Z. Lu, J. R. Shi, and S. Q. Shen, Phys. Rev. Lett. **107**, 076801 (2011).

17  Shinobu Hikami, Anatoly I. Larkin, and Yosuke Nagaoka, Prog. Theor. Phys. **63** (2), 707 (1980).

18  We estimate the contact resistance by measuring a gold wire contacted by the similar FIB fabrication. Considering the fact that our device is of good Ohmic contact and the device's resistance increases over 5 times with the decreasing temperature, the contact resistance is therefore believed to be less than 20 percent in the low-temperature measurement. Such estimation gives an error of 0.05 for the alpha value of 0.47 in the HLN fitting.

19  Z. G. Li, I. Garate, J. Pan, X. G. Wan, T. Chen, W. Ning, X. O. Zhang, F. Q. Song, Y. Z. Meng, X. C. Hong, X. F. Wang, L. Pi, X. R. Wang, B. G. Wang, S. Li, M. Reed, L. Glazman, and G. H. Wang, Phys. Rev. B **91**, 041401(R) (2015).

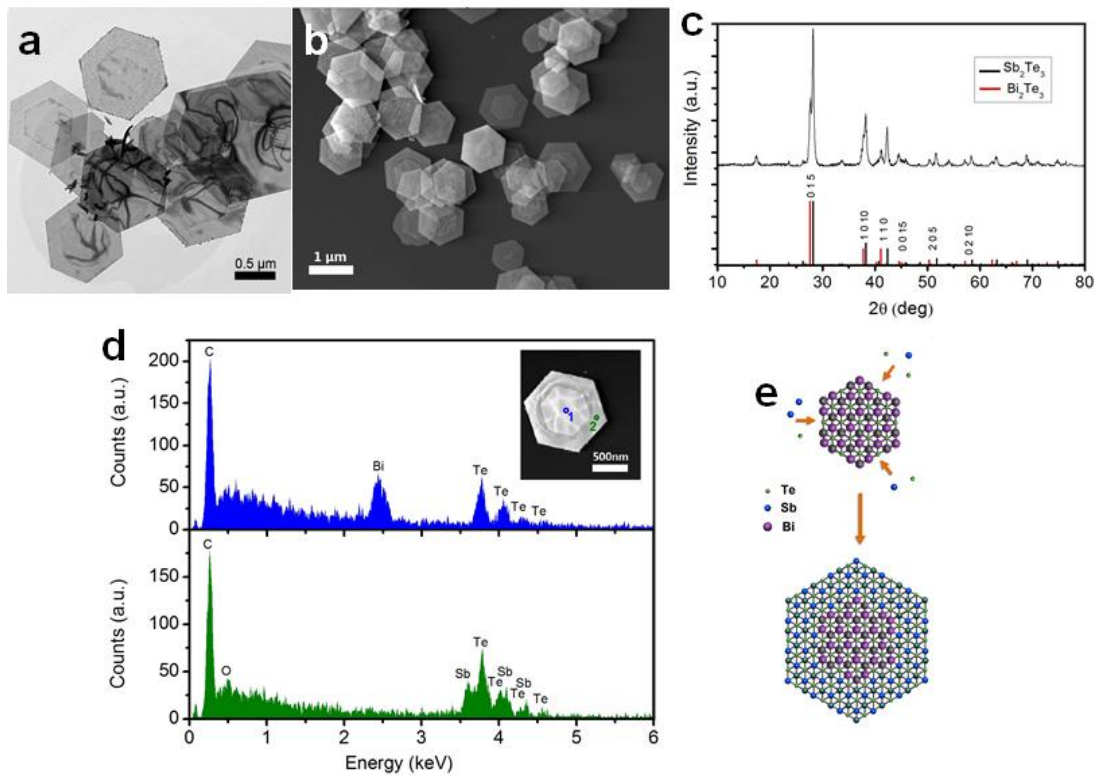

**Figure 1. Successful preparation of lateral heterojunctions of topological insulators. (a).** Transmission electron microscopy (TEM) image of the as-prepared heterojunction nanoplates. The hexagonal morphology indicates that the nanoplates

crystallized well. **(b).** Scanning TEM (STEM) image of the heterojunction nanoplates, in which the contrast of the inner region is enhanced using atomic contrast in the STEM image. **(c).** The powder XRD pattern of the products. Black and red spectrum shown in the lower panel are peaks information from standard JCPDS cards No. 15-0874 ($Sb_2Te_3$) and No. 15-0863 ($Bi_2Te_3$), respectively. **(d).** STEM image of a typical nanoplate (inset) and Corresponding EDS spectrum of the spots marked in inset respectively. **(e).** Schematic diagram of lateral epitaxial growth of the $Bi_2Te_3$-$Sb_2Te_3$ heterostructures.

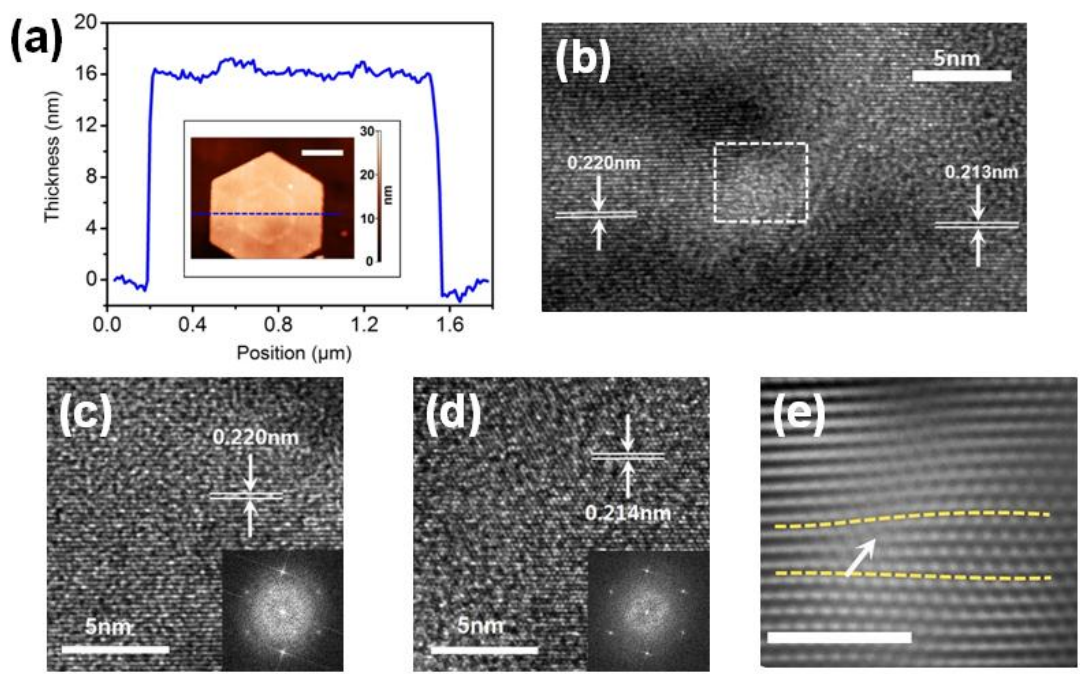

**Figure 2. Evidence for lateral epitaxial growth of the $Bi_2Te_3$/$Sb_2Te_3$ heterojunctions**. **(a).** Linear profile along the line marked in the inset (the atomic force microscopy image). The thickness was found to be 16 nm. The inner and outer crystals had the same thicknesses, excluding the possibility that core-shell structures

were formed. The scale bar in inset represents 500 nm. **(b).** High-resolution (HR) TEM image of positions near the LHJ interface. **(c and d).** HR-TEM image of the **(c)** inner $Bi_2Te_3$ and **(d)** outer $Sb_2Te_3$ zones. The insets show the fast Fourier transform patterns. The hexagonal lattice fringes showed that the lattice spacings of 0.22 and 0.214 nm applied to the (11$\bar{2}$0) planes of the crystals. **(e).** A magnified view of the region outlined by the dashed square in **(b)**. This region was the interface between the two crystalline regions. Continuous lattice fringes can be seen across the heterostructure interface. The arrow marks a dislocation and the two yellow dashed lines mark bent lattice planes, both phenomena being caused by the lattice mismatch. The scale bar represents 2 nm.

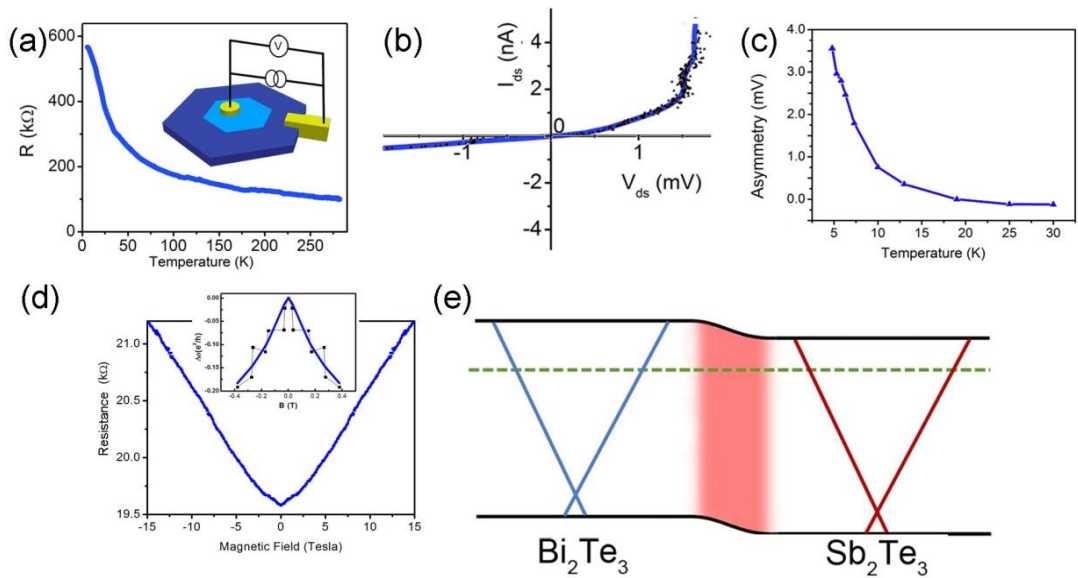

**Figure 3. Electrical transport in the $Bi_2Te_3$/$Sb_2Te_3$ heterojunction. (a).** Resistance versus temperature for a typical device. The inset shows a schematic of the device. **(b).** Current–voltage (I–V) curve of a $Bi_2Te_3$–$Sb_2Te_3$ heterojunction near zero voltage, indicating the onset of the junction field. **(c).** The temperature dependence of I–V asymmetry, defined as the mean ΔV value. ΔV is the voltage difference between corresponding positive and negative currents. **(d).** The magnetoresistance of the device up to 15T. The inset is the Hikami–Larkin–Nagaoka fitting curve for weak antilocalization near the zero field. **(e).** The proposed band diagram for the lateral heterojunction.